# Strong vortex pinning in Co-doped BaFe$_2$As$_2$ single crystal thin films

C. Tarantini[1], S. Lee[2], Y. Zhang[3], J. Jiang[1], C. W. Bark[2], J. D. Weiss[1], A. Polyanskii[1], C. T. Nelson[3], H. W. Jang[2], C. M. Folkman[2], S. H. Baek[2], X. Q. Pan[3], A. Gurevich[1], E. E. Hellstrom[1], C. B. Eom[2], D. C. Larbalestier[1]

[1] Applied Superconductivity Center, National High Magnetic Field Laboratory, Florida State University, Tallahassee, FL 32310, USA

[2] Department of Materials Science and Engineering, University of Wisconsin-Madison, Madison, WI 53706, USA

[3] Department of Materials Science and Engineering, University of Michigan, Ann Arbor, Michigan 48109, USA

We report measurements of the field and angular dependences of $J_c$ of truly epitaxial Co-doped BaFe$_2$As$_2$ thin films grown on SrTiO$_3$/(La,Sr)(Al,Ta)O$_3$ with different SrTiO$_3$ template thicknesses. The films show $J_c$ comparable to $J_c$ of single crystals and a maximum pinning force $F_p(0.6T_c) > 5$ GN/m$^3$ at $H/H_{irr} \sim 0.5$ indicative of strong vortex pinning effective up to high fields. Due to the strong correlated c-axis pinning, $J_c$ for field along the *c*-axis exceeds $J_c$ for H//*ab* plane, inverting the expectation of the $H_{c2}$ anisotropy. HRTEM reveals that the strong vortex pinning is due to a high density of nanosize columnar defects.

The discovery of superconductivity in the ferropnictides has aroused great interest due to their high critical temperature $T_c$ up to 56 K[1,2,3,4], very high upper critical fields $H_{c2}$ together with the relatively low anisotropy ($\gamma$ ~1-6) and irreversibility field $H_{irr}$ close to $H_{c2}$[5,6]. However, the presence of secondary phases at grain boundaries (GBs) in bulk materials[7,8] initially hindered the investigation of the true intragrain critical current density $J_c$. Epitaxial films allow one to understand many fundamental properties including the limits of $J_c$, giving a much better view of the true potential of ferropnictides for applications. The deposition of 1111 (REFeAsO) thin films has been particular challenging due to the difficulties of controlling the fluorine content[9], whereas in the 122 ($BaFe_2As_2$) the growth conditions can be better controlled due to the low vapour pressure of cobalt [10,11,12]. In several recent studies[11,13,14] of both 1111 and 122 pnictides the maximum self-field $J_c$ at 5 K was about 10 kA/cm$^2$, 2-3 orders of magnitude lower than in single crystal samples[6] Recently we have deposited high quality Ba-122 epitaxial thin films using a new method of template engineering, which enabled us to avoid weak linked GBs[15,16].

In this paper we study the effect of the $SrTiO_3$ (STO) template on the superconducting properties of Co-doped Ba-122 thin films, focusing on the interesting $J_c$ properties caused by strong vortex pinning. We performed transport measurements of $J_c(H)$ and its angular dependence $J_c(\theta)$, correlating them to the structural properties determined by high resolution transmission electron microscopy (HRTEM). The 350nm Co-doped Ba-122 films were grown on (001)-oriented $(La,Sr)(Al,Ta)O_3$ (LSAT) substrates using an intermediate template of 50-150 unit cells (u.c) of $SrTiO_3$. All layers were grown by pulsed laser deposition, as reported in detail elsewhere[15,16]. Figure 1 shows the superconducting transitions for which the critical temperature $T_{c,\rho=0}$ increases from 18.5 to 20.5K upon increasing the STO template thickness from 50 to 100 u.c.. A further increase of the STO thickness results in a small reduction of $T_c$. The $\theta$-$2\theta$ patterns investigated by four-circle x-ray diffraction reveal only *00l* reflections indicative of a *c*-axis growth, while the φ-scans confirm that a single in-plane orientation dominates in all films.

The field dependencies of $J_c(T,H)$ for all the films are shown in Figure 2. Both $J_c(H)$ and the irreversibility field $H_{irr}$ are enhanced by increasing the STO thickness from 50 to 100 u.c.. For instance, the self-field $J_c$ at 12 K doubles its value reaching $J_c > 0.1$ MA/cm$^2$, while $H_{irr}$(12K) determined by the Kramer extrapolation improves from about 18 T (in both field orientations) to 20-24 T (H//*c* and H//*ab*, respectively). The film on 150 u.c. STO/LSAT shows a field dependence of $J_c$ similar to the 100 u.c. film but $J_c$ is slightly suppressed by its lower T$_c$. The lower-temperature $J_c$ curves have comparatively weak field dependence up to 10-12 T due to their high $H_{irr}$ but it is even more striking that $J_c$ for H//*c* is larger than $J_c$ for H//*ab* in all samples at every temperature. Consistent with $H_{c2}$ being higher for H//*ab*, a crossover between $J_c$(H//*c*) and $J_c$(H//*ab*) is observed or estimated at the highest fields (e.g. ~15 T and 14 K or at ~17 T and 12 K in 100 u.c. STO/LSAT film), which indicates strong pinning along the c-axis. However this crossover seems to be absent for the 50 u.c. STO/LSAT film for which the irreversibility field along the *c*-axis exceeds $H_{irr}^{//ab}$. The strong c-axis pinning effectiveness is emphasized by the pinning force curves at the same reduced temperature (T ~ 0.6T$_c$) in Figure 3. The 100 u.c. STO/LSAT exhibits a remarkable $F_{p,max}$ above 5 GN/m$^3$ at about 9.5 T, which is better than the performance of the optimized NbTi at 4.2K. For the 50 u.c. STO/LSAT film the maximum pinning force $F_{p,max}$ ~3.5GN/m$^3$ occurs even at higher field ~12.5 T.

Figure 4 compares the normalized pinning forces $F_p/F_{p,max}$ as a function of the reduced magnetic field $h=H/H_{irr}$ at different temperatures for the best sample (100 u.c. STO/LSAT). For H//*ab* (Fig. 4a), the pinning force is maximum at $h_{max}$ ~ 0.2 at 16 K but $h_m$ increases at lower T. This trend suggests that pinning at higher temperatures is weakened by thermal fluctuations of vortices. At lower temperatures, the role of thermal fluctuations diminishes, and $h_{max}$ increases to ~ 0.3-0.4. For H//*c* (Fig. 4b), $F_{p,max}$ at 16K occurs already at $h_{max}$ ~ 0.4 while at lower temperatures the curves superimpose with a peak at about $h_{max} \sim 0.5$. This behaviour implies strong vortex core pinning by a dense array of nanoscale defects aligned along the c axis which reduce the effect of thermal fluctuations.

To study the mechanisms of the crossover between $J_c(H//c)$ and $J_c(H//ab)$, we also measured the angular dependences of $J_c(\theta)$ shown in Fig. 5. For all magnetic fields, $J_c(\theta)$ of the sample on 100 u.c. STO/LSAT exhibits a strong and broad *c*-axis peak. Since $H_{c2}^{//ab}$ is larger than $H_{c2}^{//c}$, such behaviour is opposite to what is expected from the mass anisotropy. At low field pinning becomes effective over the whole angular range, significantly flattening $J_c(\theta)$. At larger fields, the c-axis peak still remains pronounced but in a smaller range around 90° so the *a*-axis peak due to the mass anisotropy emerges.

The microstructural origin of strong vortex pinning was revealed by TEM images, shown in Fig.6. The planar view of the 122 film grown on 100 u.c. STO/LSAT shows a high density of randomly distributed defects with average diameter of 5 nm. The inset of Fig.6 shows they are columnar defects vertically aligned and grown from the template with only a slight meandering. From the planar view TEM we estimated a defect mean distance of 16-17 nm corresponding to a matching field of about 7-8 T comparable to the peak position of the pinning force. Such density corresponds to a volume fraction ~ 8%. This value is smaller than the critical concentration ~10% above which current blocking by pins starts decreasing $F_{p,max}$[17] so $J_c$ of our films could be further increased by incorporating more columnar pins. For H//c, $F_p(h)$ at low temperatures follows the scaling behaviour $F_p(h) \propto h^p(1-h)^q$ with $p \approx q \approx 1$ and the maximum $F_p$ at $h = H/H_{c2} \approx 0.5$, indicating strong core pinning of vortices. For H//ab, the exponents $p \approx 0.65$ and $q \approx 2.32$ at 16K and $p \approx 0.8$ and $q \approx 2.24$ at 14K seem to suggest collective pinning affected by thermal fluctuations of vortices.

The effectiveness of the templated substrate is evident by comparing $J_c(H,T,\theta)$ of our samples with other thin film reports. For instance, Sr122 thin film grown on different substrates and exhibiting good XRD patterns had $T_{c,\rho=0}$ ~15-16K, $\Delta T_c$ ~2-3K and $J_c \approx$ 10-20 kA/cm$^2$ at 4-5K [10,11,13]. By contrast we obtained $J_c$ ~0.1 MA/cm$^2$ at 12K in our best sample. Another striking feature of our films is their very strong vortex pinning for H//c. Recently Iida et al.[18] reported angular dependences and TEM of high quality Ba-122 thin film on LSAT substrates. They observed no grain boundary or other defects but, as expected in very clean samples, $J_c$ shows no c-axis peak and is low due to the lack of correlated pinning. Unlike the angular dependence data reported by Maiorov et al.[13] where a *c*-axis peak was observed only up to 1T, our data show that for H//*c,* pinning remains effective up to the maximum field measured (16T). Such strong vortex pinning by a high density of columnar defects whose matching field is close to the maximum of $F_p$(H) also reduces thermal fluctuations. Taken together, these studies suggest that vortex pinning in this 122 structure is really versatile, and also that it is important to suppress granular effects to understand the high $J_c$ of which the system is capable.

The comparison between films grown on STO templates of different thickness reveals that a further improvement of the superconducting properties is still possible. In fact, whereas the 100 u.c. STO/LSAT sample has the highest $T_c$ and largest $J_c$, the best pinning conditions are reached in the 50 u.c. STO/LSAT where the maximum of the pinning force occurs close to $H_{irr}$ ($h_m$~0.6) and $H_{irr}^{//c} > H_{irr}^{//ab}$ over the whole measured temperature and field ranges. Increasing the pin volume fraction by optimizing the 122 growth conditions without $T_c$ suppression seems quite feasible.

Work at NHMFL was supported under NSF Cooperative Agreement No. DMR-0084173, by the State of Florida, and by AFOSR grant FA9550-06-1-0474. Work at the University of Wisconsin was supported by funding from the DOE Office of Basic Energy Sciences under award number DE-FG02-06ER46327. TEM work at the University of Michigan was supported by the DOE grant DE-FG02-07ER46416. TEM facility was supported by NSF

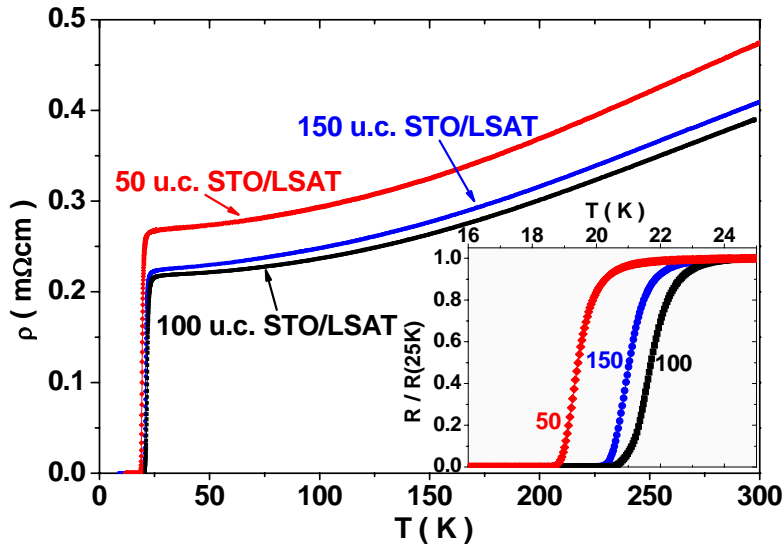

FIG.1 Resistivity versus temperature for the 3 characterized samples.

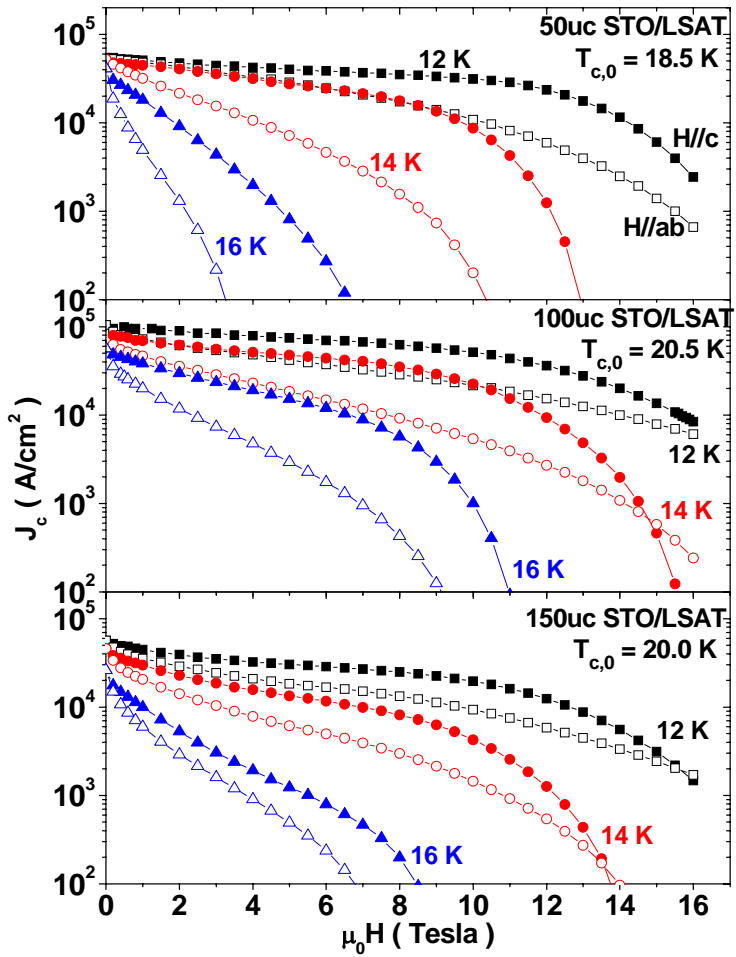

FIG.2 $J_c$ as a function of $\mu_0 H$ for the samples on 50, 100 and 150 u.c STO/LSAT. Open symbols denote H//*ab*, solid H//*c*.

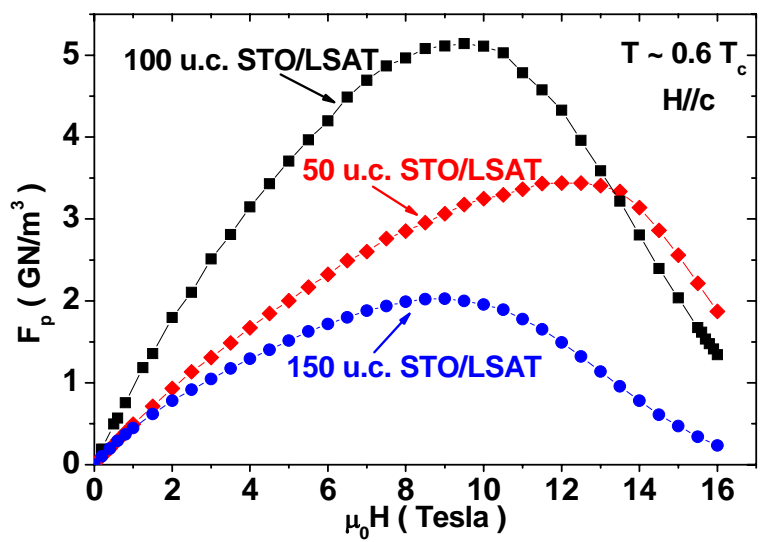

FIG.3 Pinning force as a function of field (H//*c*) at the same reduced temperature for the samples on 50, 100 and 150 u.c STO/LSAT.

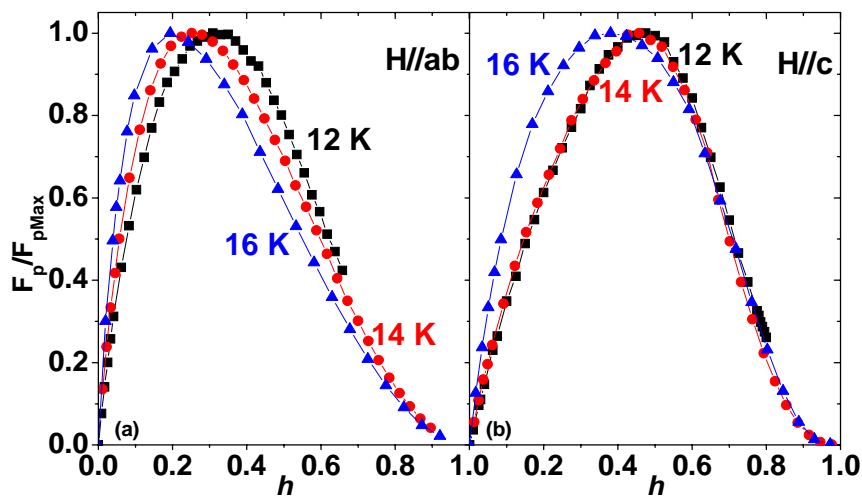

FIG.4 Normalized pinning force as a function of reduced field varying temperature for the film on 100u.c STO/LSAT.

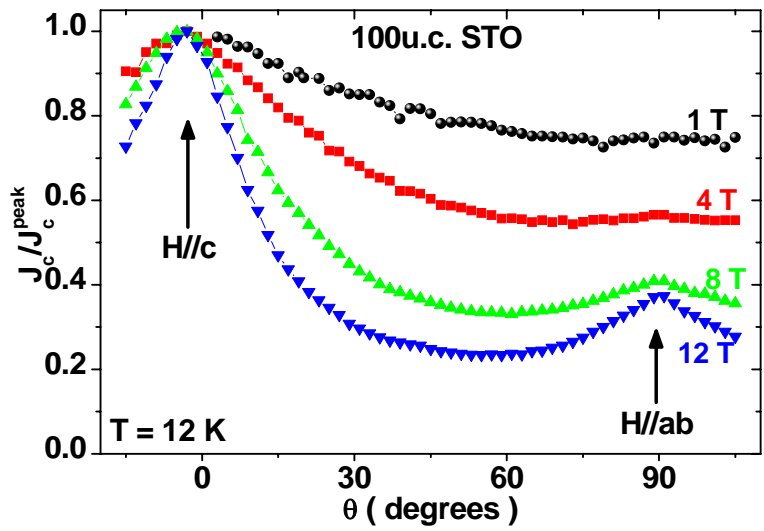

FIG.5 Normalized $J_c(\theta,12K)$ at different applied field for the film on 100u.c.STO/LSAT.

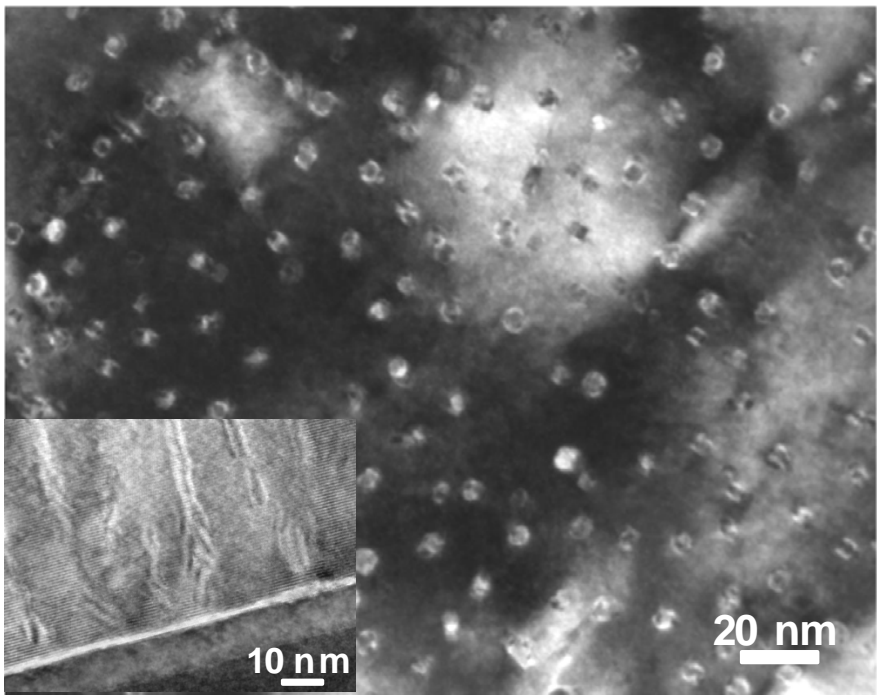

FIG.6 Planar view TEM of 100 u.c. STO/LSAT. In the inset a cross section of the 50 u.c. STO/LSAT.